\documentclass{lhep}

\journal{LHEP}
\vol{xx}
\jyear{2019}
\pages{xxx}

\received{xx January 2019}
\published{xx March 2019}

\usepackage{amsmath,amssymb,fontenc}

\usepackage{latexsym,graphicx,epsfig}

\def\lb{\linebreak[4]}
\newcommand{\be}{\begin{equation}}
\newcommand{\ee}{\end{equation}}
\newcommand{\bea}{\begin{eqnarray}}
\newcommand{\eea}{\end{eqnarray}}
\newcommand{\bes}{\begin{subequations}}
\newcommand{\ees}{\end{subequations}}
\newcommand{\bear}{\begin{equation}\begin{array}}
\newcommand{\eear}[1]{\end{array}\label{#1}\end{equation}}
\newcommand{\beg}{\begin{equation}\begin{gathered}}
\newcommand{\eeg}{\end{gathered}\end{equation}}
\newcommand{\beal}{\begin{equation}\begin{aligned}}
\newcommand{\eeal}{\end{aligned}\end{equation}}
\newcommand{\begg}{\begin{gather*}}
\newcommand{\eegg}{\end{gather*}}

\def\ba{$$\begin{array}}
 \def\ea{\end{array}$$}
\newcommand{\fr}[2]{\dfrac{{ #1}}{{ #2}}}
\newcommand{\pa}{\partial}
\newcommand{\la}{\langle}
\newcommand{\ra}{\rangle}
\newcommand{\fn}[1]{\footnote{{#1}}}
\newcommand{\bu}{$\bullet$\ }

\def\cl{\centerline}

\begin{document}

\date{\today}
\title{Problems with variable  Hilbert space in
quantum mechanics }
\author{I.~F.~Ginzburg,\\
{\it Sobolev Institute of Mathematics, Novosibirsk, 630090, Russia;\\ \it Novosibirsk State University, Novosibirsk, 630090, Russia}\\[1mm]
}


\begin{abstract}
The general problem is studied on a simple example. A quantum particle in an infinite one-dimensional well potential  is considered.
 Let the boundaries of well changes  in a finite time $T$.
The standard methods for  calculating probability of  transition from an initial to the final state  are in general inapplicable since the states of different wells belong to different Hilbert spaces.

If the final well covers only a part of the initial well (and, possibly, some outer part of the configuration space),
the total probability of the transition from any stationary state of the initial well into {\bf all} possible states of the final well is less than 1 at $T\to 0$. If the problem is regularized with a finite-height potential well, this missing probability can be understood as a non-zero probability of transitions into the continuous spectrum, despite the fact that this spectrum disappears at the removal of regularization. This phenomenon ("transition to nowhere'') can result new phenomena in some fundamental problems, in particular at description of earlier Universe.

We  discuss also how to calculate the probabilities of  discussed transitions at final $T$ for some ranges of parameters.
\end{abstract}

\maketitle

\section{Introduction}
The paper combines the solution of two problems, motivated by common source, described in the sect.~\ref{secform}.

In the sect.~\ref{secHilbgen} we discuss a new phenomenon caused by a change of the Hilbert space.

In the  sect.~\ref{secrecipe} we discuss  technical problem of useful methods of calculations in a certain range of parameters.

The discussions of obtained results are presented in each section~\ref{secHilbgen}, \ref{secrecipe} separately.

\section{Formulation of the problem}\label{secform}

Quantum particle in the infinite well $U$  is the standard entry-level problem in many quantum mechanics textbooks (see e.g. \cite{LLQM}-\cite{Gal})
 \beg
\hat{H}=\fr{\hat{p}^2}{2m} +U_i(x)\,,\\ U_i(x)=\left\{\begin{array}{ccl}
\infty: &x\in & (-\infty,0)\,,(b,\infty)\,,\\
0: &x\in & [0, b]\,.\end{array}\right.
\label{baspot}
\eeg
The stationary states of this problem are  $|n\ra_i \equiv|n;(0,b)\ra$ (the subscript $i$ means {\it initial}) with
\beg
|n\ra_i\to\psi_{n,i} =\left\{\begin{array}{cl}
 0:&x \in  (-\infty,0),\\
\sqrt{\fr{2}{b}}\sin\fr{\pi n x}{b}:& x\in [0,b],\\
 0:&x \in  (b,\infty)\,;
 \end{array}\right.
 \\ E_n=\fr{(\pi\hbar n)^2}{2mb^2},\quad n=1,\,2,\,...
\label{bassol}\eeg
These states form a basis for the Hilbert space
${\cal{H}}_i\equiv{\cal{H}}(0,b)$  of continuous square-integrable functions defined on the support $[0,\,b]$ and vanishing at its endpoints.

The stationary states for another ({\it final}) well $f$ with boundaries
\be
[0,\,b]\to [a,\,a+b\alpha]\label{newwell}
\ee
are described by equations \eqref{bassol} with the change $$x\to x-a,\quad b\to b\alpha,$$ e.~g.  $|n\ra_f\equiv |n;(a,a+b\alpha)\ra$. They form a basis for another Hilbert space ${\cal{H}}_f\equiv{\cal{H}}(a,a+b\alpha)$, defined on the support $[a,\,a+b\alpha]$. (Spaces ${\cal{H}}_f$ and ${\cal{H}}_i$ are isomorphic but do not coincide).

{\it To simplify the presentation, we set in calculations  $a=0$ almost everywhere.}

\bu \ The following problem is easily solved:
\be
\boxed{\begin{array}{c}\mbox{Let the width of  the well changes  instantly,}
\;\; b\to b\alpha.\\
\mbox{Find the probability
$W_{nk}^{if}$ of the transition}\;\; |n\ra_i\to |k\ra_f.
\end{array}}\label{jump1}
\ee

The result is obtained from eq.~\eqref{bassol}. In particular, for the  shrinking well ($\alpha<1$):
\beg
W_{nk}^{if}\equiv \left|M_{nk}^{if}\right|^2,\;\; |M_{nk}^{if}\!|=\!\left|{}_i\la n |k\ra_f\right|=
\fr{2k\sqrt{\alpha}}{\pi}\cdot\left| \fr{\sin(\pi n\alpha)}{k^2-(n\alpha)^2}\right|
\,.
\label{trampl}\eeg

\bu \  A ``natural'' modification of this problem  is
\be
\boxed{\begin{array}{c}\mbox{Find the similar probability $W_{nk}^{if}$ in  the case when}\\
\mbox{ width changes as}\;\;b\to b\alpha(t)\;
\mbox{over a finite time}\; T.\end{array}}\label{contprobl}
\ee
This problem is non-trivial. The difficulties that have arisen persist even with a more general view of  transitions at $T\to 0$.

From the formal point of view this problem is described by the Schr\"{o}dinger equation
\bes\label{HilbShred}
\beg
i\hbar\fr{d\psi(x,t)}{dt}=\hat{H}\psi(x,t)=\\=\left(-\fr{\hbar^2}{2m}\fr{d^2}{dx^2}+
U_i(x)+\hat{V}_P(x,t)\right)\psi(x,t)\,
\label{schreq}
\eeg
in which  the initial potential $U_i(x)$ \eqref{baspot}
is supplemented by perturbation $V_P(x)$
\bear{c}
\hat{V}_P(x,t)\!=\!
\ba{cc}
&\left\{\begin{array}{ccl}
0: &x\in & (-\infty,b\alpha(t))
,\\ -\infty: &x\in & (b\alpha(t),b],\\
0: &x\in & (b,\infty);\
\end{array}\right. at \;\alpha<1\,,\\[3mm] &\left\{\begin{array}{ccl}
0: &x\in & (-\infty,b)\,,\\ \infty: &x\in & (b,b\alpha(t)]\,,\\
0: &x\in & (b\alpha(t),\infty)\,.
\end{array}\right. at \;\;\alpha>1\,.
\ea\eear{pertHilb}
\ees

Standard calculations with this  perturbation are impossible for two reasons.\\
{\it (i)} Matrix elements of the  operator  $\hat V_P$ are either $\infty$ or $0$.\\
{\it (ii)} The standard form of perturbation theory   (see e. g. \cite{LLQM}) uses decomposition of time dependent wave functions in eigenfunctions of an initial  problem. However these functions belong to different Hilbert spaces  ${\cal{H}}_i$ and ${\cal{H}}_f$, making mentioned decomposition to be senseless.

Nevertheless, it is naturally to hope that there is some range of parameters (values of $\alpha$ and $T$) in which perturbation can be treated as weak one so that some form of time dependent perturbation theory  is applicable.
This situation generates two problems, discussed below.

{\bf 1. How to characterize new phenomena (if they exist) appearing when the Hilbert space changes even at $\pmb{T\to 0}$} (sect.~\ref{secHilbgen}).

{\bf 2. To present a regular method for calculation of the transition probabilities,
which  would allow one to use some  known approximate methods in the cases when it looks natural} (sect.~\ref{secrecipe}).

\section{Problems with Hilbert spaces}\label{secHilbgen}\setcounter{equation}{7}

\subsection{Regularization}\label{secreg}

It should be noted that in the calculations of transition probabilities  \eqref{trampl} we used the eigenfunctions, defined both within wells and out of wells (where these eigenfunctions are equal 0). It can be treated as the fact that we consider the  Hilbert spaces  ${\cal{H}}(0,b)$,  ${\cal{H}}(0,b\alpha(t))$ and that these spaces are subspaces of entire Hilbert space  $L_2$ of  continuous square-integrable functions defined on real axis $(-\infty,\,\infty)$, i.~e.  ${\cal H}(0,\,b),\,{\cal H}(0,\,b\alpha)\subset L_2$.

In reality, the infinite well potential is only an approximation which is useful for calculations.  The regularized problem is closer to  reality: {\it The infinite height wells are replaced by the  wells of large height $V$, which is not changed when the well width  varies}.  One considers the same Schr\"{o}dinger equation \eqref{schreq} with the following {\bf regularized} potential (for the initial well)
\be
U_{reg}^{(1)}(x)=\left\{\begin{array}{ccl}
V: &x\in & (-\infty,0)\,,\\
0: &x\in & [0, b]\,,\\
V: &x\in &(b,\infty)
\end{array}\right.\;\; \left(V\gg \fr{(\pi\hbar)^2}{2mb^2}\right). \label{regpot}
\ee
The change of the well is described by the change
 $b\to b\alpha(t)$.

At this regularization properties of $n$-th state differs weakly from those of initial well at  $$\xi_n=E_n/V= (\pi n\hbar)^2/(2mb^2 V)\ll 1.$$ At approaching $\xi_n$ to 1 the regularized picture becomes differ  from non-regularized one.

In this approach both initial and final situations are described by functions belonging to the equipped Hilbert space ${\cal {OL}}\supset L_2\supset {\cal{H}}_i, {\cal{H}}_f$. Our problem corresponds to the limit
 $V\to \infty$ (removal of regularization).

With this regularization, the standard type calculations with some improvements becomes consistent but, unfortunately, very bulky.

\subsection{Total probability of transitions}\label{secnew}

We study the essential feature of relation between the Hilbert spaces  through analysis of the instantly changing well boundary.
Let us consider the probability for the transition of some state of the  initial well $|n\ra_i$
into the any state of the final well $|k\ra_f$, forgetting the regularization. This probability is
\beg
W(n; i|f)= \sum\limits_k \left|M_{nk}^{if}\right|^2=\\=  \la n|_i\left\{ \sum\limits_k|k\ra_f\, \la k|_f\right\}|n\ra_i\equiv \la n|_i \mathbb{I}_{f} |n\ra_i \,.
\label{nkshorttot}\eeg
Here we define the operator $\mathbb{I}_{f}\equiv\sum\limits_k |k\ra_f \la k|_f$. It acts as the unit operator in the space ${\cal H}_f$. Eq.~\eqref{nkshorttot} determines, in fact, how this operator acts in the other space ${\cal H}_i$ for our problem. It can be understood by two ways, giving coinciding results. First of all, one can summarize probabilities of individual transitions \eqref{trampl}. Second, we use definition of the operator $\mathbb{I}_{f}$ as the projector to the segment $(0, b\alpha)$. (We denote $\alpha(T)=\alpha$).

\bu \ For the expanding well ($\alpha>1$) we have\\
\cl{$W(n; i|f)=\int\limits_0^b dx\; \psi_{n,i}^*(x)\psi_{n,i}(x)=1$ (norma\-li\-zation).} In other words, function $|n\ra_i$, normalized on the initial  interval, keeps the normalization in  the Hilbert state with the expanded support.

\bu \ For the shrinking well ($\alpha<1$) the initial  normalization integral lost the interval ($b\alpha\,,b$), so that
we have\lb
$W(n; i|f)=\int\limits_0^{b\alpha} dx \psi_{n,i}^*(x)\psi_{n,i}(x)<1$. The summation of   individual probabilities \eqref{trampl} naturally gives the same  result\fn{Note: if $n\alpha$ is integer, we have $\psi_{n,i}(x=b\alpha)=0$, eq.~\eqref{trampl} gives
$W_{nk}^{if}= \alpha\delta_{k,n\alpha}$
and
 $W(n; i|f)=\alpha$.}
\bear{lll}
W(n; i|f)&=&\left\{\begin{array}{l}
\int\limits_0^{b\alpha} dx\, \fr{2}{b}\sin^2\left(\fr{\pi nx}{b}\right)\\
\fr{4\alpha}{\pi^2}\sum\limits_{k=1}^\infty \fr{k^2\sin^2(\pi n\alpha)}{{(k^2-n\alpha^2)^2}}\end{array}\right\}=\\&=&\alpha\left(1-\fr{\sin(2\pi n\alpha)}{2\pi n\alpha}\right)<1\,.
\eear{totprobnarrowing}

\subsection{Disappearance of probability}\label{disc1}

According to~\eqref{totprobnarrowing}, at $\alpha<1$ some fraction of probability disappears. In the  discussed regularization picture it means that some part of the initial state goes over into the continuous spectrum, despite  the fact that this spectrum disappears when the regularization is removed.
It is clear  that this phenomenon takes place also in the more general case of moving boundaries if the final well does not cover the initial one, for example, at the shift of boundaries \eqref{newwell} with $a>0$ for both $a+b\alpha>b$ and $a+b\alpha<b$.

Therefore,  we find that {\it the description in terms of only initial and final wells appears incomplete. It should be supplemented by information about a big system embracing both these wells\fn{Certainly, one can try to implement this information  in the form of   more refined regularization procedure.}.} In the considered toy example properties of this big system are  sufficiently clear.

\subsection{Possible value of probability disappearance}

The situation with the change of Hilbert space is realized in the Nature at phase transitions.

In the standard description of phase transitions in quantum systems both initial and final Hilbert spaces are described often by corresponding sets of elementary excitations. The discussed "loss of probability" means that a complete description may require the inclusion of normally skipped  degrees of freedom of some embraced system.

For the phase transitions in the matter the properties of this big system are usually  almost evident. The most interesting problems with the change of the Hilbert space are those in which properties  of a big system are not at all obvious,one can speak that part of the initial state is lost who knows where. The example of such type transition is the cosmological  phase transition with electroweak symmetry breaking (after inflation). It is hardly possible to say with certainty know what is a large system containing both initial and final Hilbert spaces.

We find that this rearrangement of the Hilbert space can be accompanied by the loss of states. {\it If it indeed happens, what is the fate of these lost states? Can they be the source for dark energy?} In fact, similar opportunity was discussed recently with respect to "the discreteness of Universe at the Planck scale (naturally expected to arise from quantum gravity)" \cite{Sudar}.

\section{Methods for calculation}\label{secrecipe}\setcounter{equation}{10}


Now we consider more technical problems -- we present a regular methods for calculation of the transition probabilities,
which  would allow one to use some  known approximate methods in the cases when it looks natural. This problem was discussed by many authors for different particular laws of boundary motion (see e.~g.~\cite{Gal56}-\cite{Gal}, \cite{adiabappr}-\cite{lectwalls}). In this section we don't pretend for new results but present brief  review of developed methods, allowing news only in some details.

We note that the main parameter of the problem for the evolution of $n$-th state is the  time of motion of walls $T\sim 1/\alpha'_t$ in its relation to the characteristic time of life $\tau_n=\hbar/E_n$.

\subsection{Slow motion of wells. Adiabatic case}

If the walls move slowly ($T>\tau_n$), the adiabatic approximation looks reasonable \cite{Gal56}, \cite{adiabappr}. In this approximation, the evolution of the system is considered as a sequence of stationary states of the infinite well with new width. The previous state  in this sequence is the starting one for the subsequent one.  In the main approximation, the probability of $n\to n$ transition is close to 1 (even at big $\alpha$ or $1/\alpha$), other probabilities are small \cite{Gal56}.

\subsection{Fast motion of wells. 1}

If the condition $T>\tau_n$ is violated, the calculations without enormously large intermediate quantities can be based on the mapping of time dependent well to the initial well~\cite{map1},~\cite{lectwalls}. We present here some variant of this approach. At  $T<\tau_n$ this approach allows to apply  method which is similar to the standard perturbation theory.

The mapping by itself is the rescaling  of the coordinate\fn{At $a(t)\neq 0$ one can use the mapping $y=(x-a(t))/\alpha(t)$.}\lb $y=x/\alpha(t)$.
In this new variable, the potential  keeps form $U_i(y)$ \eqref{baspot}  at all times. However, the form of the kinetic term is modified, $d/dx\to \alpha^{-1}\pa/\pa y +y^{-1}\pa/\pa\alpha$. Besides, the transformation of the wave function and the scale of time are useful:
\beg
(A):\;\;y=x/\alpha(t)\,,\\ (B):\;\;\psi=\sqrt{y}\chi\,,\quad (C):\;\;
\tau=\int\limits_0^t\fr{dt}{\alpha^2(t)}\,.\label{timescale}
\eeg
Now the Schr\"{o}dinger equation is transformed to the form
\bear{c}
-i\hbar\fr{d\chi}{d\tau}=\left(\hat{H_1}+\hat{V}\right)\chi\,;\\[3mm]
\hat{H_1}=-\fr{\hbar^2}{2m}\left(\fr{\pa^2}{\pa y^2}-\fr{3}{4y^2}
\right)+U_i(y)\,,\\[3mm]
\hat{V}=-\fr{\hbar^2}{2my^2}\left(\alpha^2\fr{\pa^2}{\pa \alpha^2}
+2y\alpha\cdot \fr{\pa^2}{\pa y\pa \alpha}\right).
\eear{Schrnew}

The  eq.~(\ref{timescale} (C)) can be treated as that for obtaining dependence $\alpha$ on $\tau$. It allows to transform differentiation with respect to $\alpha$ to the differentiation with respect  to $\tau$.

We treat $\hat{H_1}$ as the new non-perturbed Hamiltonian and term  $\hat{V}$ as the perturbation.

The eigenstates $\chi_n^w(y,\tau)$ of the Hamiltonian $\hat{H_1}$ are continuous functions vanishing at the boundaries  of well (superscript $^w$  marks these states). They are expressed in terms of the Bessel function   $J_1(z)$ and the values of its zeroes $z_n$, i.~e. solutions of equation $J_1(z_n)=0$:
\beg
\chi_n^w(y,\tau)= \fr{\sqrt{2}}{b|J_2(z_n)|}
\sqrt{y}J_1(yz_n/b)e^{-iE^w_n\tau/\hbar}\,,\\ E^w_n=\fr{(\pi n\hbar )^2}{2mb^2}u_n^2\,,\quad z_n=\pi n u_n\,;
\\
\int\limits_0^b \chi_n^w(y,\tau)^*\chi_m^w(y,\tau)dy=\delta_{mn}\,.
\label{basappr}\eeg
We find useful to express  zeroes of the Bessel function in the special form $z_n=\pi n u_n$ with factor $u_n$ which is close to 1:
\begin{table}[htb]
\tbl{\it Values $u_n-1$ for different $n$.
\label{tab1}}{
\begin{tabular}{|c|c|c|c|c|c|}\hline
n&1&2&3&...&n\\\hline
$u_n-1$&0.22&0.116&0.08&...&$<(1/4n)$\\\hline
\end{tabular}}
\end{table}

\noindent This closeness  means that  eigenvalues and eigenfunctions of $\hat{H_1}$ are close to those for the initial problem \eqref{bassol}.

\subsection{Fast motion of wells. 2. Perturbation}

Note that the factor
$1/y^2$ in $\hat{V}$ do not  obstruct the convergence
of matrix elements $V_{nk}^w$ since $\chi_n(y)\to Ay^{3/2}$ at $y\to 0$.

Main features of solution are seen well in the simplest example with the linear dependence  $\alpha(t)=1+\alpha'\, t$, where\lb $\alpha^\prime=(\alpha-1)/T$. In this case the integration of \eqref{timescale} results in
 \be
\tau=\fr{1}{\alpha^\prime}\left(1-\fr{1}{\alpha}\right)\; \Rightarrow\;
\alpha=\fr{1}{1-\alpha^\prime\tau}\,,\qquad \tau
=\fr{t}{\alpha(t)}\,.\label{linpertHilb}
\ee

After that the perturbation operator is transformed to the form
\be
\hat{V}=-\fr{\hbar^2}{2my^2}\left[ \fr{1}{\alpha^2\alpha^{\prime 2}} \fr{\pa^2}{\pa\tau^2}-\fr{2}{\alpha\alpha^\prime} \fr{\pa}{\pa\tau}+\fr{2y}{\alpha\alpha^\prime}\fr{\pa^2}{\pa y\pa\tau}\right]\,.
\label{linpert}
\ee

To estimate conditions for applicability of the standard type perturbation theory to the discussed problem, we note  that in the matrix element $V_{nm}^w$ the operator $\pa/\pa\tau$ gives factor $(E_n^w-E_m^w)/\hbar$. Hence, for this transition the expansion parameter  is $\delta=(E_n^w-E_m^w)/(\hbar\alpha\alpha^\prime)\sim T/\tau_{nm}$ (at $|\alpha(\alpha-1)|\sim 1$ we have  $T\sim T$). The perturbation theory works at $\delta\ll 1$, i.~e. in the case when  a time of motion of walls, $T<\tau_{nm}$.  Note that if this condition is valid for the states with moderate $n$, it is violated for the states with very large $n$ (typical situation for validity of perturbation theory for high lying levels in quantum mechanics).

\section*{Acknowledgment}

I am thankful to A. Filippov,  D. Gorbunov, I. Ivanov, D.~Ivanov, S.~Kutateladze,  S.~Panfil, V.~Pavlov, V.~Serbo, G.~Shestakov, B.~Sikach, A.~Slavnov, V.~Tikhomirov, S.~Vodopianov for  useful discussions.
The work was supported in part by  the program of fundamental scientific researches of the SB RAS ¹ II.15.1., project ¹ 0314-2016-0021 and by
the National Science Center, Poland, through the HARMONIA project under contract UMO-2015/18/M/ST2/00518.\\

\end{document}